\documentclass[pra,twocolumn,showpacs,amssymb]{revtex4}
\usepackage{amsmath}
\usepackage{pdfpages, epsfig}

\begin{document}
\title{Quasiparticle Dispersions and Lifetimes in the Normal State of the BCS-BEC Crossover}

\author{Matthew D. \surname{Reichl}}
	%\affiliation{Laboratory of Atomic and Solid State Physics, Cornell University, Ithaca, New York 14853, USA}
	
\author{Erich J. \surname{Mueller}}
	\affiliation{Laboratory of Atomic and Solid State Physics, Cornell University, Ithaca, New York 14853, USA}

\date{\today}


\begin{abstract}        % give a summary of your paper
We compute the spectral density in the normal phase of an interacting homogenous Fermi gas using a T-matrix approximation. We fit the quasiparticle peaks of the spectral density to BCS-like dispersion relations, and extract estimates of a ``pseudo-gap" energy scale and an effective Fermi-wavevector as a function of interaction strength. We find that the effective Fermi-wavevector of the quasiparticles vanishes when the inverse scattering length exceeds some positive threshold. We also find that near unitarity the quasiparticle lifetimes, estimated from the widths of the peaks in the spectral density, approach values on the order of the inverse Fermi-energy. These results are consistent with the ``breakdown of Fermi liquid theory" observed in recent experiments.
 
%                         please supply keywords within your abstract
\end{abstract}
\maketitle 

\section{Introduction}

Interacting degenerate Fermi gases have attracted continued interest since their experimental realization over a decade ago \cite{demarco1999}. Below a critical temperature $T_c$ these gases exhibit a superfluid state which can be continuously tuned via Feshbach resonances \cite{chin2010} from a BCS state of Cooper pairs, to a BEC state of tightly bound bosonic molecules \cite{regal2007}. While the physics of the superfluid phase in this BCS-BEC crossover is well established \cite{zwerger2011}, much less is known about the normal phase at temperatures above $T_c$. 

One theme that has emerged from theoretical and experimental investigations at $T>T_c$ is the idea of a ``pseudogap" phase in the middle of the crossover, where the density of states at the Fermi energy is suppressed due to strong many-body pairing effects. If and how this phase emerges in the BEC-BCS crossover has long been a source of experimental and theoretical investigation (see Ref.~\cite{chen2014} for a comprehensive review). More recently, there has been interest \cite{gaebler2010, nascimbene2011, perali2011, sagi2014, doggen2014} in a related question: How does Fermi-liquid theory \cite{baym2008}, which is expected to be valid in in the BCS regime, break down when crossing over to the BEC regime where the normal phase is a gas of weakly interacting bosons? There has been some disagreement in the conclusions drawn from experiments which use radio-frequency (RF) spectroscopy to probe the single particle spectral density in harmonically trapped systems. The observed RF spectrum in the crossover region seems to be well described by both Fermi-liquid theory \cite{nascimbene2011} and theories displaying a pseudogap phase \cite{gaebler2010, perali2011}. A more recent experiment \cite{sagi2014} probed the spectral density in a nearly \textit{homogenous} system which avoids the density inhomogeneity of the trapped systems that can obscure features in the RF spectrum \cite{amaricci2014}. This experiment found evidence that the well defined quasi-particles one expects from Fermi-liquid theory become absent as the interactions are tuned from the BCS to the BEC side.

In this paper we address this problem theoretically by computing the single particle spectral density within a T-matrix approximation as a function of interaction strength. We find near unitarity that the quasiparticles represented by peaks in the spectral density have short lifetimes at the Fermi-wavevector. We also find that an effective Fermi-wavevector extracted from the shape of the quasiparticle dispersions vanishes when one moves sufficiently deep into the BEC regime. Both of these observations point toward a breakdown in the Fermi-liquid description of the normal phase in which there is a well-defined Fermi surface and long-lived quasiparticles at the Fermi-wavevector. Previous theoretical works using related T-matrix approximations have similarly discussed the shape and widths of peaks in the spectral density \cite{perali2002, tsuchiya2009, palestini2012}. Much of that work focused on the BEC regime \cite{perali2002} or on the temperature dependence of the spectral density at a few discrete values of the interaction strength \cite{tsuchiya2009, palestini2012}. We extend these results and systematically explore the dependence of the spectral density on interaction strength.

This paper is organized as follows. In Sec.~\ref{sectmat} we discuss the T-matrix approximation in detail. In Sec.~\ref{secresults} we show numerical results for the spectral densities and describe our procedure for analyzing the quasiparticle dispersions and lifetimes. In Sec.~\ref{secdiscussion} we discuss our results and conclude.

\begin{figure*} \vspace{1.0em}
\hbox{\hspace{3.0em}
\includegraphics[width=0.9\textwidth]{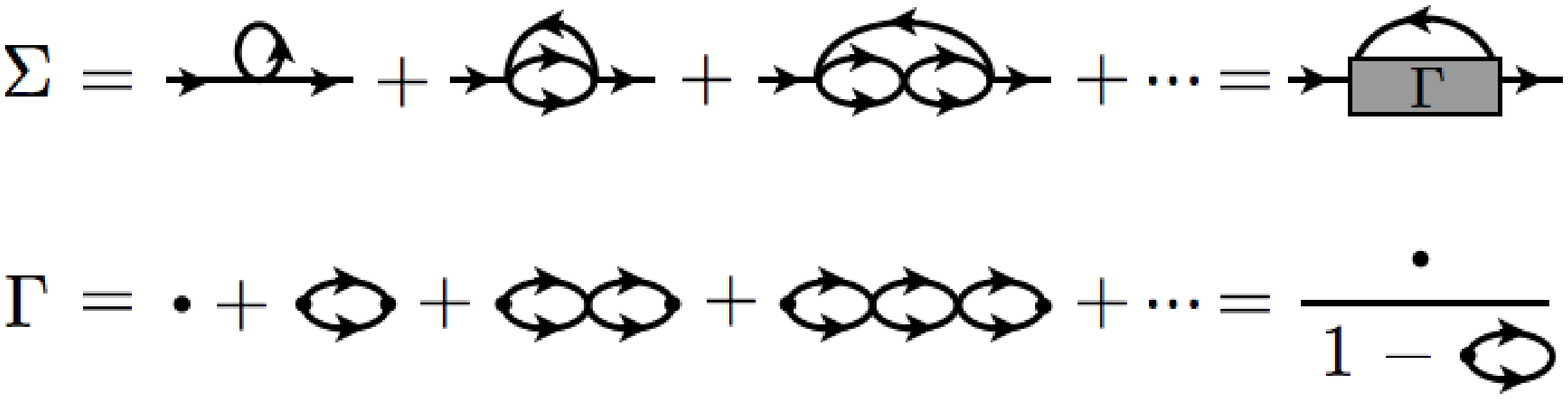}}
\caption{Diagrammatic representation of the T-matrix approximation used in this paper (Eq.~(\ref{eqtmat})).} 
 \label{figfeyndiag}
\end{figure*}

\section{T-Matrix Approximation} \label{sectmat}
In this paper we consider a Hamiltonian that models two component fermions with attractive s-wave interactions
\begin{equation}
H= \sum_{\mathbf{k}} \epsilon_{\mathbf{k} } a_{\mathbf{k}\sigma }^\dagger a_{\mathbf{k}\sigma } + \frac{g}{V} \sum_{\mathbf{k}\mathbf{p}\mathbf{q}} a^\dagger_{\mathbf{k} \uparrow}a^\dagger_{\mathbf{p} \downarrow} a_{\mathbf{p}-\mathbf{q} \downarrow}a_{\mathbf{k}+\mathbf{q} \uparrow}
\end{equation}
where $\epsilon_\mathbf{k}= \frac{k^2}{2m} -\mu$ and the interaction strength $g$ is related to the scattering length $a$ by $g^{-1}= \frac{m}{4\pi a} - \frac{1}{V} \sum_{\mathbf{k}} \frac{m}{k^2}$ (we set $\hbar=1$ and Boltzmann's constant $k_B=1$ throughout).

We use a non-self-consistent T-matrix approximation \cite{perali2002, strinati2012},  to calculate the single-particle self energy $\Sigma$:
\begin{eqnarray} \label{eqtmat}
&&\hspace{-2em}
\Sigma(\mathbf{q},i\omega_n)=  \frac{T}{V} \sum_\mathbf{k} \sum_m \Gamma(\mathbf{k}, i\Omega_m) G_0(\mathbf{q}-\mathbf{k}, i\Omega_m -i\omega_n)\nonumber\\
&&\hspace{-2em}\Gamma(\mathbf{q}, i \Omega_m)^{-1}  =  \frac{m}{4\pi a} - \frac{1}{V} \sum_{\mathbf{k}} \frac{m}{k^2} \\\nonumber
&&\qquad+\frac{T}{V}\sum_\mathbf{k} \sum_n G_0(\mathbf{q}-\mathbf{k}, i\Omega_m -i\omega_n) G_0(\mathbf{k}, i\omega_n)
\end{eqnarray}
where $\omega_n = (2n+1)\pi T$ and $\Omega_m = 2m\pi T$ (with integers $m$ and $n$) are fermionic and bosonic Matsubara frequencies at temperature $T$. $G_0(\mathbf{k}, i\omega_n)= 1/(i\omega_n - \epsilon_\mathbf{k})$ denotes the bare single particle propagator. The density is given by
\begin{equation} \label{eqn}
n \equiv  \frac{k_F^3}{3\pi^2}= 2\frac{T}{V} \sum_\mathbf{k} \sum_n G(\mathbf{k}, i \omega_n)
\end{equation}
where $G^{-1}= G_0^{-1}-\Sigma$. The critical temperature is set by a pairing instability condition (the Thouless criterion) 
\begin{equation}
\Gamma^{-1}(\mathbf{q}=0, i \Omega_m =0)|_{T=T_c}=0
\end{equation}

The expressions in Eq.~(\ref{eqtmat}) are derived by summing the infinite subset of Feynman diagrams that include scattering processes occurring in the vacuum two-body problem (this shown in Fig.~\ref{figfeyndiag}). This approximation, which neglects interactions between \textit{pairs} of fermions, nevertheless accurately models the physics in the weakly interacting BCS regime ($\frac{1}{k_Fa}<<-1$, where the normal phase is a Fermi-liquid) and in the strongly interacting BEC regime ($\frac{1}{k_Fa}>>1$, where the normal phase near $T_c$ is a gas of free bosonic molecules of mass $2m$). In particular in the deep BCS regime, the Thouless criterion is equivalent to the expression for $T_c$ from BCS theory; in the deep BEC regime, the Thouless criterion together with Eq.~(\ref{eqn}) yield the correct condensation temperature for a gas of non-interacting bosons \cite{nozieres1985}.

 In the crossover regime ($-1<\frac{1}{k_F a}<1$) this approximation is less well-controlled and there is no \textit{a priori} reason to expect it to be accurate. However previous studies using \textit{ab initio} techniques \cite{magierski2011} and self-consistent T-matrix theories \cite{haussmann2009} (where the bare propagators in Eq.~(\ref{eqtmat}) are replaced by the fully dressed propagator $G$) produce results with similar qualitative features. Since we are mainly concerned with discussing the qualitative physics within this regime, we use the relatively simple approximation described by Eq. (\ref{eqtmat}).
 
After performing the Matsubara sums in Eq.~(\ref{eqtmat}) and taking $ i \omega_n \to \lim_{\delta \to 0} \omega +i \delta$ we arrive at the following expressions for the imaginary part of the self energy:
%\begin{equation} \label{eqimsig}
%\begin{split}
%&\textrm{Im}\Sigma(\mathbf{q}, \omega)=\\
% &-\frac{1}{2} \int \frac{d^3k}{(2\pi)^3} B(\mathbf{k}+\mathbf{q}, \omega+\epsilon_{\mathbf{k}})  [f_B(\omega+\epsilon_\mathbf{k})+f_F(\epsilon_\mathbf{k})]
%\end{split}
%\end{equation}
\begin{eqnarray} \label{eqimsig}
&&\textrm{Im}\Sigma(\mathbf{q}, \omega)=\\ \nonumber
 && \quad\quad\quad
 \int \frac{d^3k}{(2\pi)^3} B(\mathbf{k}+\mathbf{q}, \omega+\epsilon_{\mathbf{k}})  [f_B(\omega+\epsilon_\mathbf{k})+f_F(\epsilon_\mathbf{k})]
\end{eqnarray}
where $f_{F}(\epsilon) = (\exp(\epsilon/T) + 1)^{-1}$, $f_{B}(\epsilon) = (\exp(\epsilon/T) - 1)^{-1}$, and 
%\begin{equation} \label{eqimgam}
%\begin{split}
%& B(\mathbf{q}, \omega) = -2 \textrm{Im} \Gamma(\mathbf{q}, \omega)  =  \\
%&-2\lim_{\delta \to 0} \textrm{Im}[ \frac{m}{4\pi a}+ \int \frac{d^3k}{(2\pi)^3} \frac{f_F(\epsilon_\mathbf{k}) + f_F(\epsilon_{\mathbf{q}-\mathbf{k}})-1}{(\omega+i\delta)-\epsilon_\mathbf{k} - \epsilon_{\mathbf{q}-\mathbf{k}}}-\frac{m}{k^2}]^{-1}
%\end{split}
%\end{equation}
\begin{eqnarray} \label{eqimgam}
&& B(\mathbf{q}, \omega) =  \textrm{Im} \Gamma(\mathbf{q}, \omega)=    \\ \nonumber
&&\textrm{Im}\left[ \frac{m}{4\pi a}+ \int \frac{d^3k}{(2\pi)^3} \frac{f_F(\epsilon_\mathbf{k}) + f_F(\epsilon_{\mathbf{q}-\mathbf{k}})-1}{(\omega+i\delta)-\epsilon_\mathbf{k} - \epsilon_{\mathbf{q}-\mathbf{k}}}-\frac{m}{k^2}\right]^{-1} \\\nonumber
%&&=2 \lim_{\delta \to 0} \textrm{Im}\frac{1}{\frac{m}{4\pi a}+ \int \frac{d^3k}{(2\pi)^3} \frac{f_F(\epsilon_\mathbf{k}) + f_F(\epsilon_{\mathbf{q}-\mathbf{k}})-1}{(\omega+i\delta)-\epsilon_\mathbf{k} - \epsilon_{\mathbf{q}-\mathbf{k}}}-\frac{m}{k^2}}
\end{eqnarray}
with $\delta\to0$.
We numerically evaluate the integrals in Eqs.~(\ref{eqimsig}) and (\ref{eqimgam}) and extract the real part of the retarded self energy using a Kramers-Kronig relation.
 The single particle spectral density $A(\mathbf{k}, \omega)$ is then given by
\begin{equation}
\begin{split}
A(\mathbf{k}, \omega)&=-2 \textrm{Im} G(\mathbf{k}, \omega)\\
&=\frac{-2 \textrm{Im}\Sigma(\mathbf{k}, \omega)}{[\omega-\epsilon_{\mathbf{k}}-\textrm{Re}\Sigma(\mathbf{k}, \omega)]^2+[\textrm{Im}\Sigma(\mathbf{k}, \omega)]^2}
\end{split}
\end{equation}

\section{Results} \label{secresults}

\subsection{Spectral Density}

\begin{figure*} \vspace{1.0em} 
\hbox{\hspace{-0.0em}
\includegraphics[width=1\textwidth]{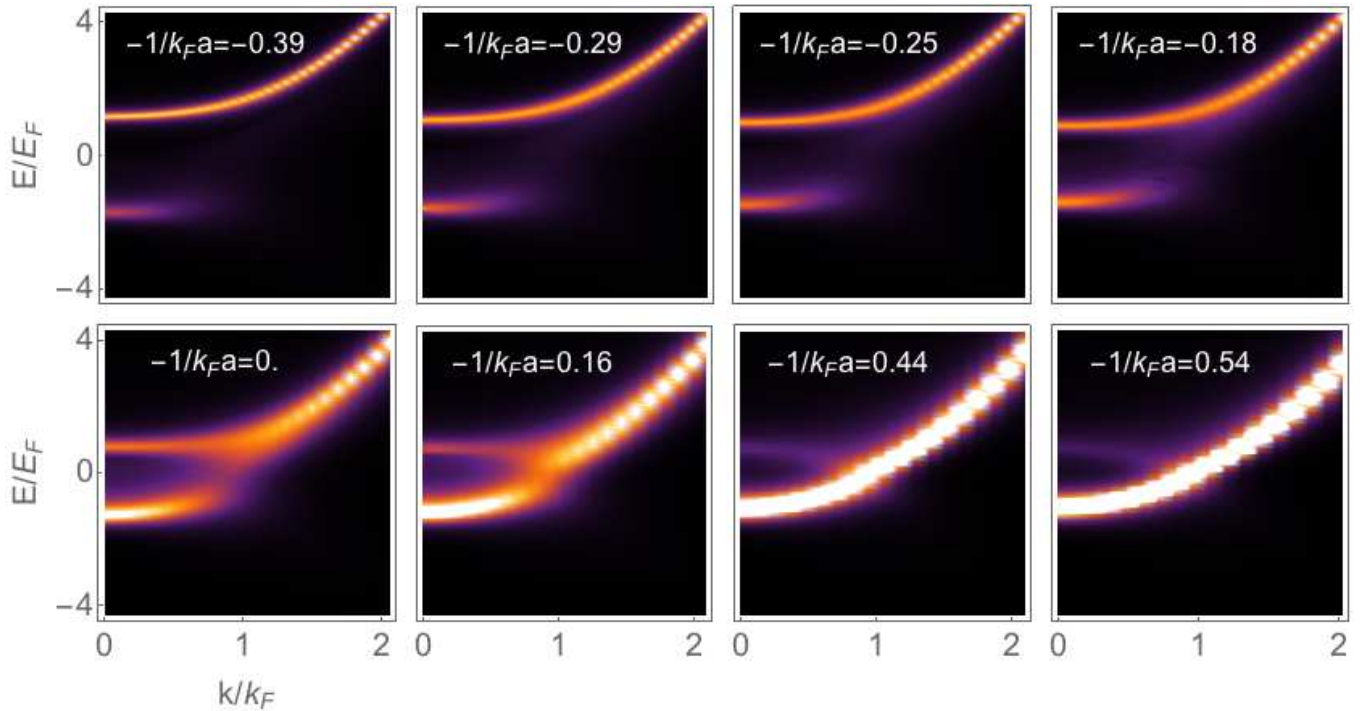}}
\caption{(Color online) Spectral density in the normal phase for various interaction strengths at temperature $T=1.1T_c$. Lighter colors correspond to higher spectral density.} 
 \label{figsdens}
\end{figure*}

Fig.~\ref{figsdens} shows the spectral density  $A(\mathbf{k}, E)$ calculated for interaction strengths $-0.5 \lesssim \frac{1}{k_F a} \lesssim 0.5$ at temperatures $T=1.1T_c$. This choice of temperature was motivated by recent experiments performed slightly above $T_c$ \cite{sagi2014}. In the weakly interacting BCS limit (for instance, the bottom right panel in Fig,~\ref{figsdens}) we observe a strong quadratically dispersing quasiparticle peak along with a faint peak at $E>0,k<k_F$. 

As the interactions become stronger we see two clearly identifiable quasiparticle dispersions, one with $E>0$ and another with $E<0$. These two branches are separated in energy by a depression (or ``pseudogap" 
\cite{tsuchiya2009, su2010, chien2010}) in the density of states centered at $E=0,k=k_F$. Finally, for very strong interactions (the top panels in Fig.~\ref{figsdens}) the two quasiparticle branches become further separated in energy and the spectral density is nearly fully suppressed near $E=0,k=k_F$.
Similar results calculated from various T-matrix approximations are shown in Refs.~\cite{tsuchiya2009, chien2010, palestini2012}.

In the BCS regime there is a simple cartoon picture \cite{mueller2011} of these results: One can add a fermion by either occupying a ``normal" fermion particle state with dispersion $E \approx \epsilon_\mathbf{k}$ or by creating a ``pair" of energy $E_{\textrm{pair}}(\mathbf{p})$ and annihilating a fermion with energy $\epsilon_{\mathbf{p}-\mathbf{k}}$. The latter excitations are broad as the pairs will be created with a range of $\mathbf{p}$. In this regime, the pair has vanishingly small energy $E_{\textrm{pair}}(0)\approx 0$, so we expect the ``pair" peak to roughly track $E\approx - \epsilon_\mathbf{k}$. As the interactions are increased, there is hybridization between these two branches that produces a psuedogap analogous to the gap that is produced by the hybridization of particle and hole states in the superfluid phase. Finally in the BEC regime, $E_{\textrm{pair}}<0$, which shifts the branch associated with pairs entirely to negative energies; the positive energy branch in this regime corresponds to adding one ``normal" fermion to a sea of paired fermions for all $k$.

\subsection{Quasiparticle Dispersions and Lifetimes} \label{secdisp}

\begin{figure} \vspace{1.0em}
\hbox{\hspace{-0.6em}
\includegraphics[width=0.45\textwidth]{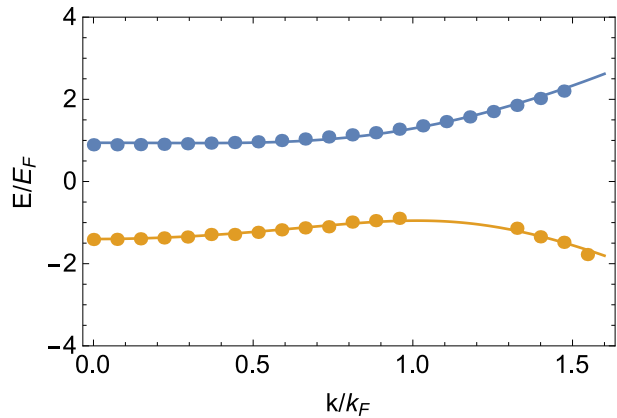}}
\caption{(Color online) Peak locations of the spectral density at $1/k_Fa=0.18$ ($T=1.1T_c$). The solid lines show least-squares fits to the dispersion relation in Eq.~(\ref{eqfit}). The board and asymmetric nature of the peaks prevents us from determining the lower branch dispersion in the range $1.0 k_F \lesssim k \lesssim1.25 k_F$.} 
\label{figdis}
\end{figure}

To describe the qualitative features of the quasiparticle dispersions, we fit the peaks in the spectral density to a phenomenological model obtained from the dispersion relations expected in the superfluid phase:
\begin{equation} \label{eqfit}
E_{\textrm{peak}}(k)_\pm= \pm \sqrt{ \frac{1}{2m}(k-k_{L_\pm})^2+\Delta_\pm^2 }
\end{equation}
where $E_{\textrm{peak}}(k)_\pm$ are the peak locations as a function of momentum for the positive and negative energy branches. $\Delta_\pm$ parametrizes the magnitude of the psuedo-gap energy scale, and $k_{L_\pm}$ parametrizes the location of the effective Fermi-wavevector. Note that $k_L$ (called the ``Luttinger-wavevector" in Refs. \cite{perali2011, palestini2012}) may not equal $k_F$ as defined in Eq.~(\ref{eqn}). An example of such a fit is shown in Fig. \ref{figdis} (where $1/k_Fa= 0.18$, $\Delta_-= 0.95 E_F$, $\Delta_+= 0.94 E_F$,  $k_{L_-}=1.01 k_F$, and $k_{L_+}=0.34 k_F$).

Fits like these have been performed in previous work in the BEC regime \cite{perali2002} and for a few select values of the interaction strength at different temperatures \cite{palestini2012}. Here we focus on the dependence of the fit parameters on interaction strength and perform fits for a number of different values of $1/k_Fa$ in the crossover regime. Similar strategies have been used to capture the qualitative features of experimental data for harmonically trapped fermions \cite{perali2011, tsuchiya2011}.

\begin{figure} \vspace{1.0em}
\hbox{\hspace{-0.6em}
\includegraphics[width=0.45\textwidth]{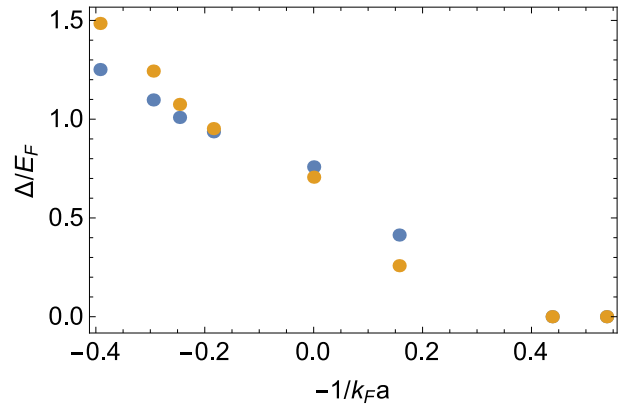}}
\caption{(Color online) Pseudogap energy scale $\Delta_-$ (orange) and $\Delta_+$ (blue) extracted from fits of Eq.~(\ref{eqfit}) to the peak locations in the spectral density ($T=1.1T_c$).} 
\label{figgap}
\end{figure}

\begin{figure} \vspace{1.0em}
\hbox{\hspace{-0.6em}
\includegraphics[width=0.45\textwidth]{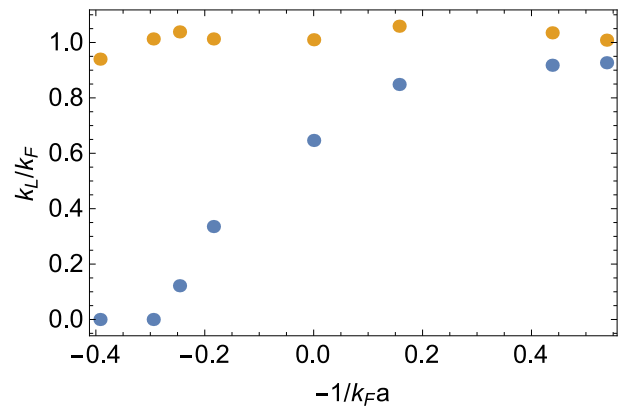}}
\caption{(Color online) Effective Fermi-wave vectors $k_{L_-}$ (orange) and $k_{L_+}$ (blue) extracted from fits of Eq.~(\ref{eqfit}) to the peak locations in the spectral density ($T=1.1T_c$).} 
\label{figk}
\end{figure}

Figs.~\ref{figgap} and \ref{figk} show the fitting parameters $\Delta_\pm$  and $k_{L_\pm}$, respectively, as a function of interaction strength. Each point was exacted by performing a least squares minimization of the difference between Eq.~(\ref{eqfit}) and the location of the quasiparticle peaks in the spectral densities in Fig.~\ref{figsdens}. The psuedogap scale is maximal on the BEC side of resonance, smoothly dropping as one approaches the BCS side. These parameters are determined by a global fit, so they are sensitive not only to features near the Fermi wave-vector, but to all $k$. While $k_{L_+}$ and $k_{L_-}$ are generically quite different, we find $\Delta_+\approx \Delta_-$.

\begin{figure} \vspace{1.0em}
\hbox{\hspace{-0.6em}
\includegraphics[width=0.45\textwidth]{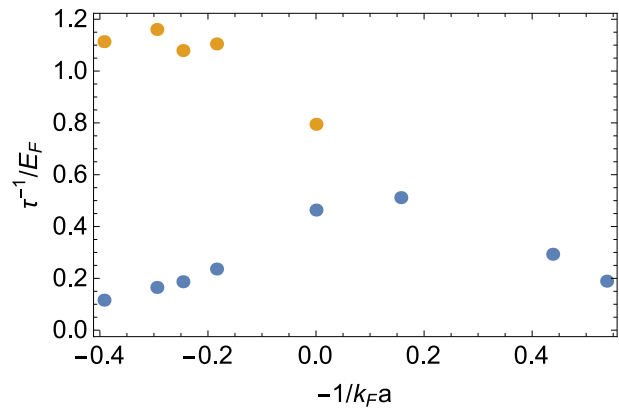}}
\caption{(Color online) Inverse quasi particle lifetimes $\tau^{-1}$ as a function of interaction strength ($T=1.1T_c$). The orange (blue) points are the spectral widths of the negative (positive) energy branch of the quasiparticle spectral peak at $k=k_F$. } 
\label{figlifetime}
\end{figure}

\begin{figure} \vspace{1.0em}
\hbox{\hspace{-0.6em}
\includegraphics[width=0.45\textwidth]{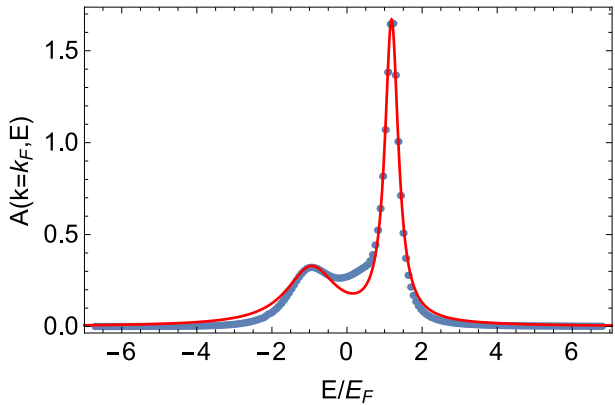}}
\caption{(Color online) $A(k=k_F, E)$ at $\frac{1}{k_Fa}=0.18$ ($T=1.1T_c$). The solid red line is a fit of the data to the sum of two Lorenzians. The two widths extracted from these fits provide estimates of the inverse quasi-particle lifetimes $\tau^{-1}$.  $\tau^{-1}$ at other interaction strengths are plotted in Fig.~\ref{figlifetime}. } 
\label{figlfit}
\end{figure}

To estimate the lifetimes of the quasiparticles we extract the widths of the spectral peaks at $k=k_F$ by fitting the peaks in the function $A(k=k_F, E)$ to Lorentzian functions. Fig.~\ref{figlfit} shows an example of such a fit for $\frac{1}{k_F a}= 0.18$. Fig.~\ref{figlifetime} shows the quasiparticle widths $\tau^{-1}$ as a function of interaction strength. We note that for $\frac{1}{k_Fa}<0$ the negative energy branch cannot clearly be separated from the positive energy branch at $k=k_F$; for these interaction strengths we only plot one value of the quasiparticle lifetime.

\section{Discussion and Conclusions} \label{secdiscussion}
Our results demonstrate a breakdown of Fermi-liquid theory in two key respects. First, as shown in Fig.~\ref{figlifetime}, the inverse lifetimes of the quasi-particles at $k_F$ increase as one approaches unitarity from the BCS-side and eventually reaches a maximum of value on the order of $E_F$. Fermi-liquid theory is predicated on a vanishingly small inverse lifetime at the Fermi-surface. This breakdown is due to both the high temperature and the strong interactions. 

Second, the effective Fermi-wavevector $k_L$, which helps identify the presence of a remnant Fermi surface \cite{perali2011, palestini2012}, vanishes for the upper branch at $\frac{1}{k_Fa} \approx 0.3$ (see Fig.~\ref{figk}). We note that the effective Fermi-wavevector for the bottom branch ($k_{L_-}$) remains fixed at $k_F$ for the range of interaction strengths shown in Fig.~\ref{figk}). However, other studies have shown that at higher interaction strengths ($\frac{1}{k_F a}\approx 0.7$), $k_{L_-}$ vanishes as well \cite{perali2011}. $k_{L_-}$ and $k_{L_+}$ are associated with different branches of excitations and therefore have distinct physical interpretations. $k_{L_+}$ is associated with particle excitations, while $k_{L_-}$ is associated with holes. On the BEC side of resonance, one can add a fermion without disturbing the pairs, but adding a hole requires breaking pairs. Thus many-body effects are less important for particles, and $k_{L_+}$ vanishes in a regime where $k_{L_-}$ is still essentially equal to $k_F$.

%At strong interaction strengths, $k_{L_+}$ is associated with the remnant Fermi surface for excitations which involve adding one fermion to the sea of fermion pairs. $k_{L_-}$ is associated with the remnant Fermi surface of the bulk equilibrium system. In the deep BEC regime all fermions are tightly bound in pairs and there is no Fermi surface for either branch.%

A recent experiment \cite{sagi2014} used RF spectroscopy to probe the spectral density of a nearly homogenous Fermi gas in the normal phase as a function of interaction strength. The RF signal reported in this work came primarily from the negative energy branch of the single particle excitations since any signal from positive energy excitations is suppressed by a Fermi factor. The authors fit their data to a two-mode model with a narrow quadratically dispersing peak expected from a Fermi-liquid theory (with weight $Z\leq 1$) and a broad ``incoherent background" (with weight $(1-Z)$) corresponding to the spectral weight of weakly interacting bosonic molecules in the normal phase. They find that the best fit to their data at interaction strengths $\frac{1}{k_fa}\gtrsim0.3$ has $Z\approx0$. The authors conclude that this is a signal of a breakdown in the Fermi-liquid description of their data.

 We caution however that because their two mode fit was performed on a single broad dispersing peak (coming from the negative energy branch of excitations), it is somewhat difficult to interpret the meaning of this vanishing Fermi-liquid contribution. Moreover, good fits to the data came at the expense of some unrealistic values of the fitting parameters. In particular, the parameter corresponding to the temperature of the incoherent bosonic contribution had best fit values nearly four times greater than the estimated temperature of the gas. 
 
One possible interpretation of this two-mode fitting procedure is that $Z$ roughly measures the weight of the component of the RF signal which disperses as $E \sim k^2$. Assuming that the peaks of the RF data are well described by the BCS-like dispersion $E_{\textrm{peak}}(k)_-$ given in Eq.~(\ref{eqfit}), the quadratically dispersing component vanishes when $k_{L_-} \to 0$. The vanishing of $Z$ in the two-mode model then seems to roughly correspond to the vanishing of $k_{L_-}$ that occurs as one approaches the BEC regime. We recommend fitting the peaks of the RF data to the BCS dispersion relations in Eq.~(\ref{eqfit}) to directly extract $k_{L_-}$ and to compare the experimental results with our T-matrix calculations.

%A recent experiment \cite{sagi2014} probed the spectral density of a nearly homogenous Fermi gas in the normal phase as a function of interaction strength. The authors of that work fit their data to a model that included the narrow quadratically dispersing peak at positive energies expected from a Fermi-liquid theory and a broad ``incoherent" peak at negative energies corresponding to the spectral weight of weakly interacting bosonic molecules in the normal phase. They find that the best fit to their data at interaction strengths $\frac{1}{k_fa}\gtrsim0.3$ has a vanishing contribution from the Fermi-liquid part of the model. Curiously, this value at which they claim a breakdown of Fermi-liquid theory is consistent with the interaction strength where we predict a vanishing effective Fermi wave vector. To more directly compare their results with our T-matrix calculations, one could fit the peaks of their data to the BCS dispersion relations that we used in Sec.~\ref{secdisp}.%

Given the high temperatures and strong interactions in the normal state near unitarity, it is not surprising that many of the features expected of a Fermi liquid are absent. It appears that the T-matrix approximation captures this physics. It remains to be seen if an alternative framework can replace these Fermi-liquid ideas. The most tantalizing steps in that direction come from exploring universal bounds on transport coefficients \cite{adams2012}.

\section{Acknowledgements}
We thank Giancarlo Strinati for comments regarding the experiment in Ref.~\cite{sagi2014}.
We acknowledge support from ARO-MURI Non-equilibrium Many-body Dynamics grant (W911NF-14-1-0003). This material is also based upon work supported by the National Science Foundation Graduate Research Fellowship under Grant No. DGE-1144153.

\bibliography{becbcs}

\end{document}